\documentclass[10pt,aps,prc,twocolumn,superscriptaddress,floatfix,nofootinbib]{revtex4-1}
\usepackage{bm}
\usepackage{amsmath,amssymb,mathrsfs}
\usepackage{geometry}
\usepackage{here}

\usepackage{graphicx}

\def\fs#1{{\footnotesize #1}}

\makeatletter
\newcommand\footnoteref[1]{\protected@xdef\@thefnmark{\ref{#1}}\@footnotemark}
\makeatother

\begin{document}

\title{
Enhanced nucleon transfer in tip collisions of $^{238}$U+$^{124}$Sn
}

\author{Kazuyuki Sekizawa}
\email[Present address: Department of Physics, University of Washington; Electronic address: ]{sekizawa@uw.edu}
\affiliation{Faculty of Physics, Warsaw University of Technology, Ulica Koszykowa 75, 00-662 Warsaw, Poland}

\date{October 19, 2017}

\begin{abstract}
Multinucleon transfer processes in low-energy heavy ion reactions have attracted increasing
interest in recent years aiming at production of new neutron-rich isotopes. Clearly, it is an
imperative task to further develop understanding of underlying reaction mechanisms to lead
experiments to success. In this paper, from systematic time-dependent Hartree-Fock calculations
for the $^{238}$U+$^{124}$Sn reaction, it is demonstrated that transfer dynamics depend
strongly on the orientations of $^{238}$U, quantum shells, and collision energies. Two important
conclusions are obtained: (i) Experimentally observed many-proton transfer from $^{238}$U to
$^{124}$Sn can be explained by a multinucleon transfer mechanism governed by enhanced neck
evolution in tip collisions; (ii) Novel reaction dynamics are observed in tip collisions at energies
substantially above the Coulomb barrier, where a number of nucleons are transferred from $^{124}$Sn
to $^{238}$U, producing transuranium nuclei as primary reaction products, that could be a
means to synthesize superheavy nuclei. Both results indicate the importance of the neck
(shape) evolution dynamics, which are sensitive to the orientations, shell effects and collision
energies, for exploring possible pathways to produce new unstable nuclei.
\end{abstract}


\maketitle

\textit{Introduction.}
The necking is one of the fundamental degrees of freedom in nuclear dynamics.
When a nucleus splits into two---the nuclear fission \cite{Bohr(1939),Hill(1953)}---the
ways of evolving a neck characterize the fission outcomes such as kinetic and excitation
energies as well as mass and charge of the fission products \cite{Schunck(2016)}.
Since the neck formation lowers the Coulomb barrier height \cite{Umar(2006),Washiyama(2008),Simenel(2013)2},
it significantly affects the fusion cross section. Moreover, the neck plays an important role for,
\textit{e.g.}, nucleon exchanges and energy dissipation \cite{Volkov(1977),Feldmeier(1987),
Ayik(2009),Washiyama(2009)1,Yilmaz(2011),Yilmaz(2014),Ayik(2015),Ayik(2016),Ayik(2017)}.
This work strengthens the importance of neck evolution dynamics in multinucleon transfer
processes that could be a key element toward synthesis of yet-unknown superheavy nuclei.

\begin{figure*}[t]
   \begin{center}
   \includegraphics[width=16.5cm]{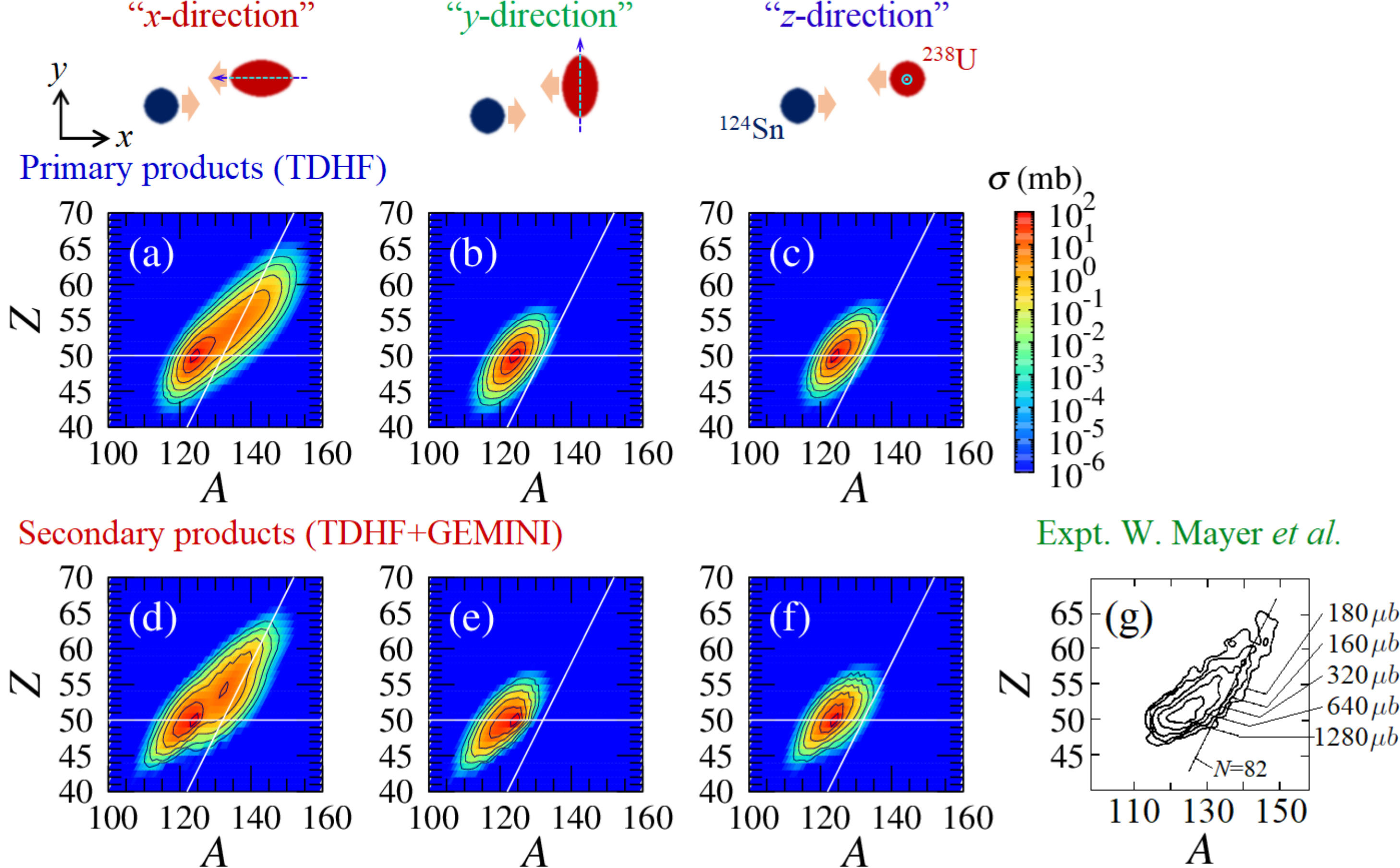}
   \end{center}\vspace{-5mm}
   \caption{(Color online)
   A comparison of production cross sections for the lighter fragments in the
   $^{238}$U+$^{124}$Sn reaction at $E_{\rm c.m.}$\,$\simeq465$~MeV.
   (a-c): Cross sections for \textit{primary} reaction products from TDHF.
   (d-f): Cross sections for \textit{secondary} reaction products from TDHF+GEMINI.
   (g): The experimental data (for reaction products with energy losses $\ge25$~MeV) reported in Ref.~\cite{Mayer(1985)}.
   The magic numbers, $Z=50$ and $N=82$, are indicated by solid lines. The contour lines for the
   theoretical results (a-f) correspond to 0.001, 0.01, 0.1, 1, 10, and 100~mb. At the top part of the figure,
   the collision geometries examined are depicted, showing the $x$-direction case [for (a,\,d)],
   the $y$-direction case [for (b,\,e)], and the $z$-direction case [for (c,\,f)]. The figure
   shown in (g) was taken from Ref.~\cite{Mayer(1985)} with permission.
   }\vspace{-3mm}
   \label{FIG:NTCS}
\end{figure*}

Recently, the multinucleon transfer reaction is considered as a promising means to produce
new neutron-rich heavy nuclei and has been extensively studied \cite{Loveland(2011),Kozulin(2012),
Kratz(2013),Kratz(2015),Barrett(2015),Vogt(2015),Barrett(2015),Watanabe(2015),Welsh(2017),
Dasso(1994),Dasso(1995),Yanez(2015),Zagrebaev(2005),Zagrebaev(2006),Zagrebaev(2007)1,
Zagrebaev(2007)2,Zagrebaev(2008)1,Zagrebaev(2008)2,Zagrebaev(2011),Zagrebaev(2013),
Zagrebaev(2014),Zagrebaev(2015),Karpov(2017),Adamian(2005),Penionzhkevich(2005),Penionzhkevich(2006),
Feng(2009),Adamian(2010)1,Adamian(2010)2,Adamian(2010)3,Mun(2014),Mun(2015),Zhu(2015),
Zhu(2016),Feng(2017),Zhu(2017)1,Li(2017),Zhu(2017)2,Zhao(2015),Li(2016),Wang(2016),Zhao(2016),
Souliotis(2003),Souliotis(2011)CoMD,Fountas(2014)CoMD}. In this context, among the pioneering
experiments \cite{Hulet(1977),Hildenbrand(1977),Schadel(1978),Glassel(1979),Essel(1979),
Freiesleben(1979),Schadel(1982),Moody(1986),Welch(1987)}, Mayer \textit{et al.}~at GSI
reported~\cite{Mayer(1985)} measurements of production cross sections for lighter (target-like)
fragments in $^{238}$U-induced dissipative collisions with $^{110}$Pd and $^{124}$Sn targets,
employing the inverse kinematics. It was observed that, for the $^{238}$U+$^{124}$Sn reaction
at $E_{\rm c.m.}$\,$\simeq465$~MeV, up to around 15 protons are transferred from $^{238}$U
to $^{124}$Sn, whereas the neutron number of the lighter fragments tends to be close to the neutron
magic number, $N=82$ [see Fig.~\ref{FIG:NTCS}(g) for the experimental data]. Similar shell effects
were observed also for the $^{238}$U+$^{110}$Pd reaction. The authors of Ref.~\cite{Mayer(1985)}
thus concluded that strong structural effects may be present in the $^{238}$U-induced dissipative
collisions, where the shell effects of $N=82$ play a crucial role during the multi-proton transfer processes.
Even though the finding is fascinating, clear theoretical explanation for this particular observation
has not yet been given so far.

In this Rapid Communication, it is demonstrated, based on microscopic time-dependent Hartree-Fock
(TDHF) calculations, that the observed multi-proton transfer processes can be explained by characteristic
neck evolution dynamics in tip collisions. Only in such a nuclear orientation, a thick and long neck is
developed in the course of the collision, and its subsequent rupture gives rise to the transfer of both
neutrons and protons from $^{238}$U to $^{124}$Sn. Because of the dissipative character of the
reaction, the reaction products are highly excited and secondary deexcitation processes affect significantly
the production cross sections. It is shown that, after the secondary particle evaporation, the neutron
number of the lighter fragments tends to be along with the magic number, $N=82$, explaining the
experimental observation. To gain deeper insight into the reaction mechanism, collision energy
dependence is also investigated for tip and side collisions, revealing qualitative difference. In this paper,
the importance of the neck evolution dynamics in low-energy heavy ion reactions is highlighted.

\textit{Method.}
In this work, the TDHF theory is employed to unveil the mechanism of multinucleon transfer
processes in the $^{238}$U+$^{124}$Sn reaction. The theory is able to describe important
dynamics during the collision such as shape deformation of the composite system, nucleon exchanges,
energy dissipation, shell effects, and so forth, without adjustable parameters. With the aid of the
particle-number projection method \cite{Projection}, one can compute production cross sections
for \textit{primary} (excited) reaction products from the TDHF wave functions after collision \cite{KS_KY_MNT}.
Very recently, a method, called TDHF+GEMINI, was proposed \cite{KS_GEMINI}, which combines the
TDHF theory with a statistical compound-nucleus deexcitation model, \fs{GEMINI}++ \cite{GEMINI++},
that allows the evaluation of production cross sections for \textit{secondary} reaction products after possible
particle evaporation and/or fission. Those methods are used to make a comparison with the experimental data.
(See, \textit{e.g.}, Refs.~\cite{Negele(review),Simenel(review),Nakatsukasa(PTEP),Sky3D,Nakatsukasa(review)},
for reviews, and Refs.~\cite{Ayik(2009),Washiyama(2009)1,Yilmaz(2011),Yilmaz(2014),Ayik(2015),Ayik(2016),
Ayik(2017),Umar(2006),Washiyama(2008),Projection,Guo(2007),Simenel(2007),Guo(2008),Umar(2008),Golabek(2009),
Washiyama(2009)2,Umar(2009),Kedziora(2010),Umar(2010),Iwata(2010),Oberacker(2010),Yilmaz(2011),
Iwata(2011),Evers(2011),Umar(2012)1,Oberacker(2012),Simenel(2012),Keser(2012),Umar(2012)2,Loebl(2012),
Fracasso(2012),Iwata(2013),Pardi(2013),Oberacker(2013),Simenel(2013)1,Simenel(2013)2,Umar(2014),Simenel(2014),
Dai(2014)1,Steinbach(2014),Dai(2014)2,Wakhle(2014),Oberacker(2014),Schuetrumpf(2014),Schuetrumpf(2015),
Umar(2015)1,Hammerton(2015),Washiyama(2015),Umar(2015)2,Umar(2015)3,Goddard(2015),Goddard(2016),
Bourgin(2016),Reinhard(2016),Schuetrumpf(2016),Stevenson(2016),Wang(2016),Umar(2016)1,Vo-Phuoc(2016),
Umar(2016)2,Godbey(2017),Singh(2017),Stone(2017),Yu(2017),Umar(2017),Schuetrumpf(2017)},
for recent applications of the TDHF theory.)

The TDHF calculations were performed using a parallel computational code \cite{MyPhD},
which has been successfully applied for various systems \cite{KS_KY_MNT,KS_KY_PNP,MyPhD,
Bidyut(2015),KS_KY_Ni-U,KS_SH_Kazimierz,KS_GEMINI,Bidyut(2017)}. For the energy
density functional, the Skyrme SLy5 parameter set \cite{Chabanat} was used. Static
calculations were performed with a box of $(24\,{\rm fm})^3$ with an 0.8-fm mesh.
The Hartree-Fock ground state of $^{124}$Sn is of oblate shape with $\beta\simeq0.11$
\cite{KS_KY_MNT}, while that of $^{238}$U is of prolate shape with $\beta\simeq0.27$
\cite{KS_KY_Ni-U}. The TDHF calculations were performed with a 3D box of $56\,{\rm fm}
\times56\,{\rm fm}\times24\,{\rm fm}$ for non-central collisions ($b\le10$~fm), while that
was $72\,{\rm fm}\times32\,{\rm fm}\times24\,{\rm fm}$ for head-on collisions. Since
$^{238}$U exhibits a large prolate deformation, the calculations were performed taking three
initial orientations of $^{238}$U: the symmetry axis of $^{238}$U is set parallel to the collision
axis ($x$ axis), set parallel to the impact parameter vector ($y$ axis), and set perpendicular to
the reaction plane ($xy$ plane); while the symmetry axis of $^{124}$Sn is always set perpendicular
to the reaction plane. Those orientations will be called $x$-, $y$-, and $z$-direction cases, respectively,
and are illustrated in the top part of Fig.~\ref{FIG:NTCS}. The same orientations were investigated
for the $^{64}$Ni+$^{238}$U reaction in Ref.~\cite{KS_KY_Ni-U}. The initial separation distance
was set to 24~fm along the collision axis. Because of the excessively large total number of protons
($Z$\,=\,142), fusion is no longer possible and always binary reaction products were observed.
The time evolution was continued until the relative distance between the two fragments exceeds 28~fm.

\textit{The origin of many-proton transfer.}
Let us begin with clarifying the origin of the experimentally observed many-proton
transfer in the $^{238}$U+$^{124}$Sn reaction at $E_{\rm c.m.}$\,$\simeq465$~MeV.
Figure~\ref{FIG:NTCS} exhibits the production cross sections for the lighter fragments
in the $A$-$Z$ plane. In the upper row (a-c), the cross sections for \textit{primary}
reaction products obtained from the TDHF calculations are shown; while, in the lower
row (d-f), the cross sections for \textit{secondary} reaction products from TDHF+GEMINI
are shown. For TDHF+GEMINI, a simplified treatment that utilizes average excitation energy
and angular momentum \cite{KS_GEMINI} was used, assuming that all the excitation energy
evaluated from the TDHF wave function after collision gets thermalized forming a compound
nucleus. Since a proper orientation average requires numerous computational effort, it has
not been achieved and, instead, the contributions from the $x\mbox{-,}$ $y\mbox{-,}$ and
$z$-direction cases are separately shown in (a,\,d), (b,\,e), and (c,\,f), respectively. The
magic numbers, $Z=50$ and $N=82$, are indicated by solid lines. In (g), the measured
cross sections reported in Ref.~\cite{Mayer(1985)} are presented. 

Let us first look at the experimental data shown in Fig.~\ref{FIG:NTCS}(g). The cross
sections take the maximum value at around the initial mass and charge numbers of the target,
$A=124$ and $Z=50$, as expected. As can be seen from the figure, the measured cross
sections extend toward the right-top part in the $A$-$Z$ plane, the direction increasing
the mass and charge of the lighter fragments, meaning that many nucleons are transferred
from $^{238}$U to $^{124}$Sn. One can also find that the neutron number of the lighter
fragments tends to be along with the magic number, $N=82$. The authors of Ref.~\cite{Mayer(1985)}
therefore conjectured that this is a multi-proton transfer process from $^{238}$U to $^{124}$Sn,
under strong influence of the $N=82$ shell closure.

Let us now turn to the theoretical results shown in (a-c) for primary products and (d-f) for
secondary products. From the figure, one can clearly see dramatic orientation dependence.
Namely, when the symmetry axis of $^{238}$U is set parallel to the collision axis [the $x$-direction
case shown in (a,\,d)], the production cross sections extend widely in the $A$-$Z$ plane.
In contrast, when the symmetry axis of $^{238}$U is set perpendicular to the collision axis
[the $y$- and $z$-direction cases shown in (b,\,e) and (c,\,f), respectively], the cross sections
distribute only narrowly around $A=124$ and $Z=50$. The important fact is that the cross sections
for the many-nucleon transfer from $^{238}$U to $^{124}$Sn remain substantially large even
after the secondary deexcitation processes, as shown in (d). Moreover, after the deexcitation
processes, the cross sections look aligned along the neutron magic number, $N=82$, consistent
with the experimental observation.

\begin{figure}[H]
   \begin{center}
   \includegraphics[width=7.32cm]{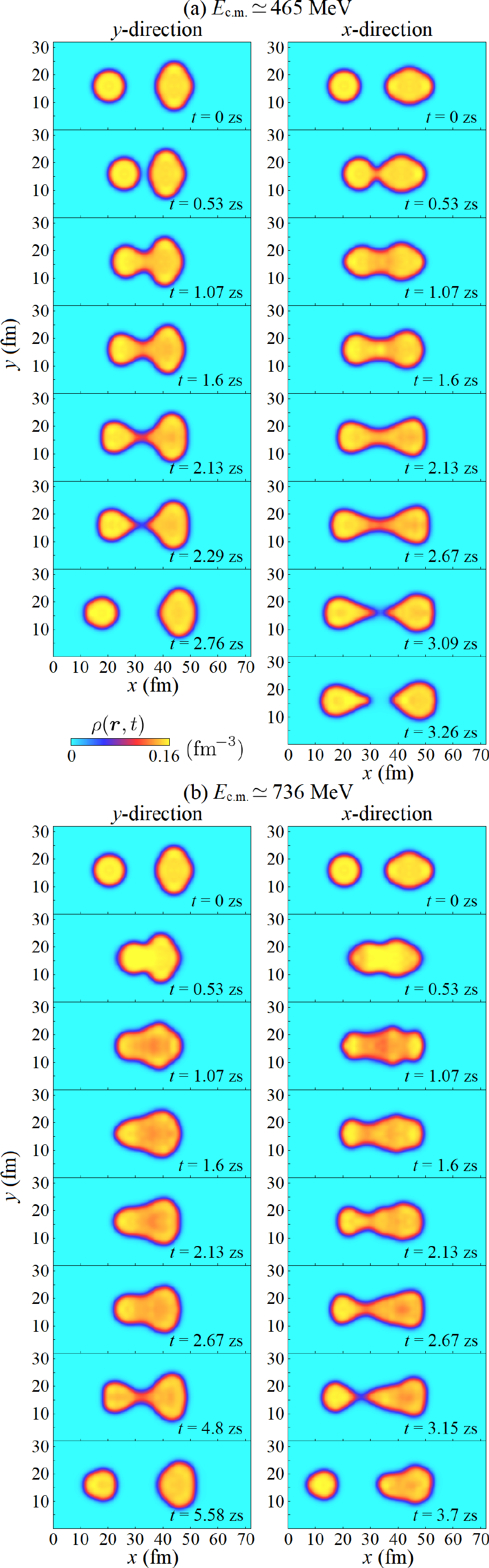}
   \end{center}\vspace{-7mm}
   \caption{(Color online)
   Snapshots of the density in $^{238}$U+$^{124}$Sn head-on collisions at
   $E_{\rm c.m.}$\,$\simeq465$~MeV (a) and 736~MeV (b).
   }
   \label{FIG:rho(t)}
\end{figure}

Why does the amount of nucleon transfer depend so much on the orientation of $^{238}$U?---the
answer lies in the remarkable difference of the neck evolution dynamics. In Fig.~\ref{FIG:rho(t)}(a),
snapshots of the density of the colliding nuclei at various times in head-on collisions of $^{238}$U+$^{124}$Sn
at $E_{\rm c.m.}$\,$\simeq465$~MeV are displayed, as an illustrative example. Time evolves from top to
bottom rows, as indicated in each panel in zeptoseconds ($1\;\mbox{zs}=10^{-21}\;\mbox{sec}$).
In the left column, the result for the $y$-direction case (side collision) is shown; while the $x$-direction
case (tip collision) is shown in the right column. From Fig.~\ref{FIG:rho(t)}(a), one can clearly see
that, when $^{238}$U collides from its tip on $^{124}$Sn (right panels), two nuclei collide deeply ($t=1.07$~zs)
and then an elongated dinuclear system is formed, evolving a thick neck structure ($t=1.6$--2.67~zs).
Since the neck ruptures at a position closer to the heavier subsystem (incident $^{238}$U in the right side),
a number of nucleons inside the neck are subsequently absorbed by the smaller fragment ($t=3.09$--3.26~zs).
Similar dynamics have been observed also for non-central collisions ($b\lesssim3$~fm). On the other hand,
when $^{238}$U collides from its tip on $^{124}$Sn (left panels), such a long neck is not developed
($t=1.6$--2.29~zs) and only few nucleons are transferred on average. It is to mention that the frozen Hartree-Fock
treatment \cite{Denisov(2002),Simenel(2017)} offers an estimate of the Coulomb barrier height, which is
$V_{\rm B}^{\rm tip}$\,$\simeq410$~MeV (\textit{i.e.}, $E_{\rm c.m.}/V_{\rm B}^{\rm tip}$\,$\simeq$\,1.13)
and $V_{\rm B}^{\rm side}$\,$\simeq448$~MeV (\textit{i.e.}, $E_{\rm c.m.}/V_{\rm B}^{\rm side}$\,$\simeq$1.04)
for the tip and side collisions, respectively, for the present system.

Summarizing, the present TDHF calculations and the analysis by TDHF+GEMINI indicate
that what was observed experimentally is the tip-collision-induced many-nucleon
transfer, which is induced by dynamics of a thick and long neck formation and breaking,
followed by secondary evaporation processes.

\textit{Energy dependence of the reaction dynamics.}
One may rise a question about the energy dependence of the neck evolution dynamics.
Namely, one may naively expect that, even in the side collision, similar multinucleon
transfer processes via the elongated dinuclear system formation and its subsequent
rupture may emerge at higher collision energies. In what follows, it is shown that
it is not the case.

\begin{figure}[t]
   \begin{center}
   \includegraphics[width=8.6cm]{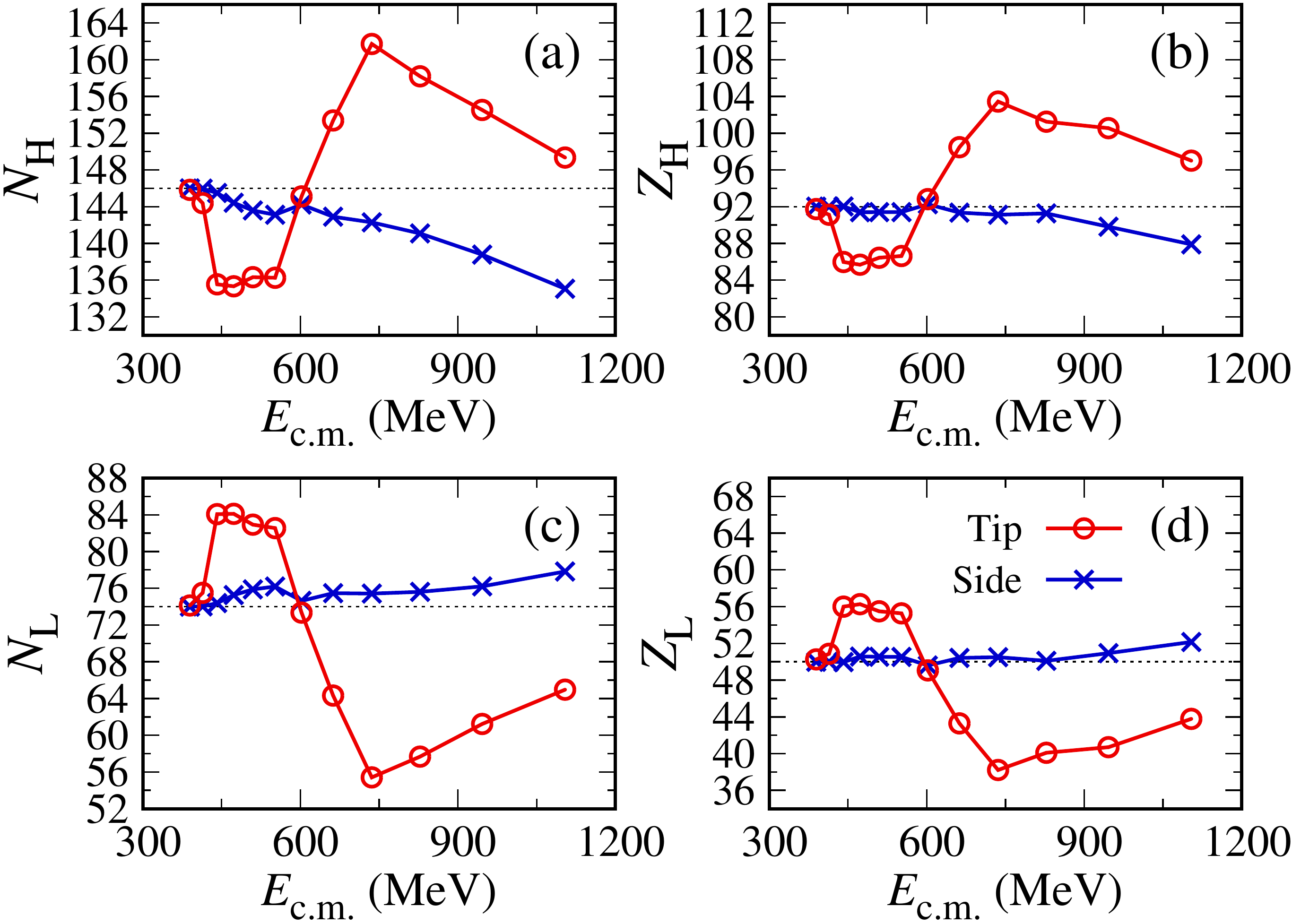}
   \end{center}\vspace{-4mm}
   \caption{(Color online)
   The TDHF results for head-on collisions of $^{238}$U+$^{124}$Sn at various center-of-mass
   energies, $E_{\rm c.m.}$. Average numbers of neutrons and protons of the heavier fragments
   ($N_{\rm H}$ and $Z_{\rm H}$) are shown in (a) and (b), respectively, while those of the lighter
   fragments ($N_{\rm L}$ and $Z_{\rm L}$) are shown in (c) and (d), respectively. The results
   associated with tip (side) collisions are shown by read open circles (blue crosses). The neutron
   and proton numbers of the projectile and the target are indicated by horizontal dotted lines.
   The frozen Hartree-Fock treatment provides the Coulomb barrier height of $V_{\rm B}^{\rm tip}
   $\,$\simeq410$~MeV and $V_{\rm B}^{\rm side}$\,$\simeq448$~MeV for the tip and side collisions,
   respectively, for this system.
   }\vspace{-2mm}
   \label{FIG:Nave_PT(E)}
\end{figure}

Figure~\ref{FIG:Nave_PT(E)} shows average numbers of nucleons of the heavier fragments (a,\,b)
and the lighter fragments (c,\,d) as a function of the center-of-mass energy. Here only head-on
collisions are investigated, taking two initial orientations of $^{238}$U, the $x$- and $y$-direction cases,
which, respectively, correspond to the tip and side collisions. From the figure, one can see that, in the
side collisions (blue crosses), the average number of nucleons of the fragments does not depend much
on the collision energy. The only visible trend is a gradual decrease (increase) of the average number
of nucleons in the heavier (lighter) fragment. The larger decrease seen in (a) as compared to the increase
seen in (c) indicates that substantial prompt neutron emissions from $^{238}$U took place during the
collision at higher energies. In the left column of Fig.~\ref{FIG:rho(t)}(b), an example of the reaction
dynamics in the side collision at $E_{\rm c.m.}$\,$\simeq736$~MeV is shown. Nevertheless two nuclei
collide so deeply and once form a compact shape without clear dinuclear structure ($t=0.53$--2.13~zs),
the system undergoes similar scission dynamics (2.67--5.58~zs), as was observed for lower energies
[\textit{cf.} the left column of Fig.~\ref{FIG:rho(t)}(a)]. The results clearly indicate that the elongated
neck structure is difficult to be developed on the equatorial side of $^{238}$U, even at higher
energies substantially above the Coulomb barrier.

In stark contrast, in the tip collisions (read open circles), dramatic collision energy dependence is
observed. Namely, at energy slightly above the Coulomb barrier ($E_{\rm c.m.}$\,$\simeq$\,442~MeV),
the average number of nucleons shows a sudden jump, which corresponds to transfer of about 10 neutrons
and 6 protons from $^{238}$U to $^{124}$Sn on average. Then it exhibits a prominent plateau pattern
in the figure around ($N_{\rm H}\simeq136$, $Z_{\rm H}\simeq86$) and ($N_{\rm L}\simeq84$,
$Z_{\rm L}\simeq56$) over a wide energy range of $442$~MeV $\lesssim E_{\rm c.m.}\lesssim 552$~MeV.
The collision energy of $E_{\rm c.m.}\simeq465$~MeV that was investigated in the previous section
actually belongs to this energy range. The latter process may be deemed as neck evolution dynamics
under the influence of the quantum shells around $N=82$ for the lighter fragment [\textit{cf.}~Fig.~\ref{FIG:Nave_PT(E)}(c)]
and $Z=82$ for the heavier fragment [\textit{cf.}~Fig.~\ref{FIG:Nave_PT(E)}(b)], although the values
do not coincide exactly with those magic numbers. It is to mention that in the plateau region the dynamics
look similar to the ones shown in the right column of Fig.~\ref{FIG:rho(t)}(a).

This is not the end of the story: \textit{i.e.}, as the collision energy increases further ($E_{\rm c.m.}
\gtrsim 552$~MeV), the plateau actually vanishes and even the direction of nucleon transfer reverses,
resulting in many-nucleon transfer from light to heavy nuclei, which may be regarded as an \textit{inverse}
(anti-symmetrizing) quasifission process \cite{Loveland(2011),Zagrebaev(2006),Zagrebaev(2007)1,
Zagrebaev(2007)2,Zagrebaev(2008)2,Zagrebaev(2011),Zagrebaev(2013),Zagrebaev(2015),Kedziora(2010)}.
At maximum, transfer of 16 neutrons and 11 protons from $^{124}$Sn to $^{238}$U is observed at
around $E_{\rm c.m.}\simeq736$~MeV. The average primary reaction products correspond roughly
to $_{38}^{93}$Sr$_{55}^{}$ and $_{103}^{265}$Lr$_{162}^{}$. The typical reaction dynamics
of the latter process are displayed in the right column of Fig.~\ref{FIG:rho(t)}(b). Since two nuclei collide
so deeply, complex surface vibration modes are induced ($t=1.07$~zs). As time evolves ($t=1.6$--2.67~zs),
a neck starts developing at a position closer to the smaller subsystem (incident $^{124}$Sn in the left side),
and eventually ruptures ($t=3.15$~zs), producing a compact lighter fragment and a strongly-deformed
heaver fragment. It seems that there is complex interplay between density fluctuations, surface vibrations,
and structural effects, \textit{e.g.}, probably shell effects around $Z=40$, in the observed inverse quasifission
process. It might also be related to dynamic clustering phenomena which were recently investigated in
light systems within the TDHF approach \cite{Schuetrumpf(2017)}. To provide a conclusive explanation,
however, further investigations are necessary, \textit{e.g.}, systematic calculations for other projectile-target
combinations at a range of collision energies, along with investigations of structural properties of the composite
system.

It should be noted that the observed inverse quasifission dynamics are different from those reported
in, \textit{e.g.}, Ref.~\cite{Zagrebaev(2013)}, where strong shell effects of $^{208}$Pb induce nucleon
transfer from $^{238}$U to $^{248}$Cm, and Ref.~\cite{Kedziora(2010)}, where a ``tip-on-side"
configuration allows nucleon transfer from the tip of $^{232}$Th to the side of $^{250}$Cf. It would be
interesting to explore similar inverse quasifission processes in, \textit{e.g.}, the $_{64}^{160}$Gd$_{96}^{}
$+$_{96}^{248}$Cm$_{152}^{}$ reaction, where shell effects of $Z$\,=\,50 (and possibly $N$\,=\,82)
or even $Z$\,=\,40, as was observed in the present system, may induce production of superheavy nuclei;
\textit{e.g.}, $_{64}$Gd\,+\,$_{96}$Cm\,$\rightarrow$\,$_{40}$Zr\,+\,$_{120}$Ubn. If one could
take advantage of shell effects around ($Z=114$ and $N=184$) for heavier fragments and ($Z=50$ and
$N=82$) for lighter fragments, the system may be able to access to the island of stability: \textit{e.g.}
$_{74}^{186}$W$_{112}^{}$+$_{96}^{248}$Cm$_{152}^{}\rightarrow_{56}^{136}$Ba$_{80}
$+$_{114}^{298}$Fl$_{184}^{}$. Of course, one has to carefully investigate the survival probability of
the primary reaction products. One should also note that possible effects of two-body dissipations may
present in collisions well above the Coulomb barrier, which need to be addressed by, \textit{e.g.},
time-dependent density matrix (TDDM) \cite{Tohyama(2016)} or molecular dynamics approaches
(see, \textit{e.g.}, \cite{Li(2016),Wang(2016),Zhao(2016),Souliotis(2011)CoMD,Fountas(2014)CoMD,
Aichelin(1991)QMD,Hartnack(1998)QMD,Feldmeier(2000)FMD,Kanada-En'yo(2012)AMD,Giuliani(2014)},
and references therein). Nevertheless, the inverse quasifission process, assisted with the expected large
(co)variances of fragment mass and charge distributions in such damped collisions \cite{Simenel(review),
Ayik(2015),Ayik(2017)}, may be a possible way to produce yet-unknown superheavy nuclei.

\textit{Conclusions.}
Production of new neutron-rich heavy and superheavy isotopes is one of the hot topics in the
nuclear physics community. In this paper, the reaction mechanism of the $^{238}$U+$^{124}$Sn
reaction has been investigated based on the microscopic framework of the time-dependent
Hartree-Fock (TDHF) theory. From the systematic TDHF calculations for the reaction at various initial
conditions, it has been demonstrated that the dynamics of neck formation and breaking, which in
turn govern the amount and the direction of nucleon transfer, depend strongly on collision energy,
quantum shells, and nuclear orientations. When $^{238}$U collides from its tip on $^{124}$Sn,
a thick and long neck is developed and a number of nucleons inside the neck are transferred when
it ruptures; whereas the neck formation is substantially hindered when $^{238}$U collides
from its side. The results have clearly shown that the experimentally observed many-proton
transfer from $^{238}$U to $^{124}$Sn, whose mechanism was a mystery for over 30 years,
may most likely be associated with the neck evolution dynamics in the tip collisions, followed
by secondary evaporation processes. Moreover, at energies substantially above the Coulomb
barrier, the emergence of novel reaction dynamics has been observed, where transuranium
nuclei are produced as a result of many-nucleon transfer from $^{124}$Sn to $^{238}$U.
The latter dynamics may be useful to create unknown superheavy nuclei. Both results strongly
suggest that the neck evolution dynamics are vital degrees of freedom that should be appropriately
taken into account in the reaction models for multinucleon transfer and quasifission processes
at low energies around the Coulomb barrier. It is to mention that some symptom of proton-pair
transfer in the $^{238}$U+$^{110}$Pd and $^{238}$U+$^{124}$Sn reactions was reported
in Ref.~\cite{Wagner(1987)}, which can be addressed by extending the theoretical framework
to include the pairing correlations \cite{Scamps(2012),Scamps(2013),Ebata(2014),Ebata(2015),
Hashimoto(2016),Scamps(2017),MSW(2017),SMW(2017),SWM(2017),Bulgac(2017)}. Lastly,
it should be underlined that the TDHF approach can predict novel reaction dynamics
in a non-empirical way, as demonstrated in this work. Therefore, further systematic TDHF
calculations for various projectile-target combinations and collision energies have potential
to open new ways to reach neutron-rich heavy and superheavy nuclei that have never
been produced to date.

\textit{Acknowledgments.}
The author wishes to thank Piotr Magierski (Warsaw University of Technology) for valuable comments
on this article. The author acknowledges support of Polish National Science Centre (NCN) Grant, decision
No.~DEC-2013/08/A/ST3/00708. This work used computational resources of the HPCI system
(HITACHI SR16000/M1) provided by Information Initiative Center (IIC), Hokkaido University,
through the HPCI System Research Projects (Project IDs: hp140010 and hp170007).
The author is grateful to Elsevier Ltd.~for the permission to reuse the figure with modification.

\end{document}